\documentclass{elsart}
\usepackage{graphicx}

\begin{document} 
\newcommand{\degg}{\mbox{$^\circ$}}
\newcommand{\ltsim}{\mbox{\tiny$\stackrel{\textstyle<}{\sim}$}}
\newcommand{\gtrsim}{\mbox{\tiny$\stackrel{\textstyle>}{\sim}$}}

\begin{frontmatter}
\title{Time resolution of Burle 85001 micro-channel 
plate photo-multipliers in comparison with Hamamatsu R2083.}
\author[KNU]{V.~Baturin},
\author[JLAB]{V.~Burkert},
\author[KNU]{W.~Kim},
\author[JLAB]{S.~Majewsky},
\author[KNU]{D.~Nekrasov},
\author[KNU]{K.~Park},
\author[JLAB]{V.~Popov},
\author[JLAB]{E.S.~Smith},
\author[KNU]{D.~Son},
\author[KNU]{S.S.~Stepanyan},
\author[JLAB]{C.~Zorn}.
%\corauth[CORRESPOND]{Corresponding authors: E.S.Smith~(elton@jlab.org) and V.Baturin~(baturin@jlab.org)} 
\corauth[CORRESPOND]{Corresponding author: Wooyoung Kim (wooyoung@jlab.org).}
\address[KNU]{Kyungpook National University, Daegu 702-701, Republic of  Korea.}
\address[JLAB]{Thomas Jefferson National Accelerator Facility, Newport News, Virginia 23606,USA}
\setcounter{footnote}{0}

%\author[JLAB]{Place.Holder for Authors}
%\address[JLAB]{Thomas Jefferson National Accelerator Facility, Newport News, Virginia 23606}

\begin{abstract}
The CLAS detector will require improvements in its particle identification 
system to take advantage of the higher energies provided by the Jefferson
Laboratory accelerator upgrade to 12 GeV. To this end, we have
studied the timing characteristics of the  micro-channel 
plate photo-multiplier $85001$ from Burle, which can be operated in
a high magnetic field environment. For reference and comparison, measurements were
also made using the standard PMT $R2083$ from Hamamatsu using two timing methods.
%been operating in coincidence. 
%Each  counter is instrumented with two  $R2083$~PMTs from Hamamtsu.
The cosmic ray method, which  utilizes
%6% identical PMTs $R2083$ from Hamamtsu and 
three identical scintillating counters $2\times3\times50~cm^3$ with PMs at the ends,
yields  $\sigma_{R2083}=59.1\pm0.7~ps$. 
%For comparison we have  measured the timing resolution 
%for the micro-channel plate PMs $Burle-85001$.
%The  method of measuring 
 The location method of  particles  from
radiative source with known coordiantes has been used to compare timing resolutions of $R2083$ and $Burle-85001$.
This  ``coordinate method''
requires only one counter instrumented with two PMs 
and it yields $\sigma_{R2083}=59.5\pm0.7~ps$ 
in agreement with the cosmic ray method.
%We extrapolate the energy dependent resolution to the M.I.P.(minimum-ionizing-particle) energy.
For the micro-channel plate photomultiplier from Burle with an
external amplification of 10 to the signals,
 the coordinate method yields $\sigma_{85001}=130\pm4~ps$.
 This method also makes it possible to   estimate  the number of primary photo-electrons.
%\newline
%\newline
%\newline
%\newline
\end{abstract}
\end{frontmatter}
%\newpage  
%\tableofcontents
PACS:          
\newpage
\section{Introduction}

The CEBAF electron accelerator at Jefferson Lab in Newport News, Virginia, 
is dedicated to exploring the nature of stronly interacting matter.
The Department of Energy has recently affirmed the decision to double
the current energy of the machine to 12 GeV by giving CD-0 approval to
the 12 GeV upgrade project\cite{upgrade}. The increased energy reach of the machine
will open up new opportunities in hadronic physics, as well as new challenges.
The experimental equipment will be upgraded to take advantage of the
higher beam energies, since both the momenta and multiplicity of secondary particles
will be significantly higher.  Therefore the particle identification
criteria and space-time resolution, as well, has to be improved.

The current time-of-flight (TOF) system\cite{r2} is the primary tool for
hadron identification in the CLAS detector. The upgraded CLAS uses 
Cerenkov light as well as TOF for hadron identification. 
%The conceptual design of the upgraded CLAS is described in \cite{upgrade}.
% shown in Fig.~\ref{clas}-a.
The new TOF
system\cite{upgrade} will have a refurbished forward-angle detector, and 
a barrel scintillation detector 
for triggering and time-of-flight
measurements in the central region.clas
%~(Fig.~\ref{clas}).
The present work concentrates on
studies which are of special interest to the barrel detector.

The nominal barrel geometry consists of 50 scintillating ``time-zero''
counters each 50~cm long and $2\times3~cm^2$ in cross section.
These counters will be placed inside the superconducting 
solenoid at a $25~cm$ radius from the beam axis. 
One of the goals of the CLAS upgrade program for the
barrel counters is to achieve a timing resolution $\sigma_{TOF}\approx 50~ps$, 
which will allow the separation of pions from kaons up to 0.64~GeV/c and pions 
from protons up to 1.25~GeV/c.

The TOF counters of CLAS\cite{r2} have two PMTs, one at each end of 
scintillation counters. Therefore, assuming the PMTs have the same resolution,
one can determine the TOF resolution   
between two  counters  as $\sigma_{TOF}$=$\sigma_{PMT}$.
However, if the TOF is measured relative to the precise   
radio frequency (RF) signal of the accelerator, 
then $\sigma_{TOF}=\frac{1}{\sqrt{2}} \sigma_{PMT}$ since the RF jitter may be neglected.
Thus we aim  to construct  a  time-zero counter with  the effective  resolution 
$\sigma_{PMT}\leq\sqrt{2}\times 50~ps$. 
We  refer to  $\sigma_{PMT}$   as  an ``effective resolution'' in order
to  emphasize that  this value  is defined  not only by the excellent  
characteristics of PMTs we use in our tests, but by the experimental environment, as well.
In particular, the barrel structure of $50$
time-zero scintillators 
%(Fig.~\ref{clas}) 
is expected to be  placed  
in the area of a high magnetic field $B\approx 2~T$, and high counting rate. Both of these can
influence the photomultiplier signals and deteriorate the  timing resolution.
If ordinary dynode photomultiplier tubes (PMTs) are used, then the only option is to 
move the PMTs out of  the region of high magnetic field, which requires bent 
light guides that are at least a meter long to transport the scintillator light to reduced field locations.
 
An alternative solution  would be the implementation of
micro-channel plate (MCP) PMs,
which were first tested as components of a scintillation time-of-flight system\cite{giles} with 
resolution of  $\approx 400ps$.
Due to  obvious  immunity of MCPs to  magnetic fields\cite{r3}, which was verified up to $2T$,
micro-channel plate PMs could be attached directly to scintillators.
Since the MCP transition time  is  short, the single electron resolution 
of MCP may in principle be as low as $\approx30~ps$\cite{r3}.
Unfortunately the resolution of scintillation  counters is  mostly dictated   
by  statistical fluctuations in the number of photo-electrons, and therefore
reduced quantum efficiency will have a direct adverse effect on timing.

In this  paper we evaluate the possibilities of achieving the goals for
the TOF barrel for the upgraded CLAS12 detector.
To this end we have conducted tests using cosmic rays and a $^{90}$Sr 
radioactive
source to study the resolution of scintillation counters (Bicron BC-408) with 
two PMTs on each.
Various tests required up to three identical counters, each
$2\times3\times50cm^3$ in size.
First we instrumented the setup with standard fast $R2083$ from Hamamatsu.
The time resolution of $R2083$ PMTs was determined with a standard cosmic-ray setup
using three identical counters. The radiative source and one of these counters was used to check the measurements with
the ``coordinate method'' described below. 
Then we replaced the fast R2083 PMTs 
by  MCP~PMs ``Burle-85001'' in the test counter and repeated the study with the ionization source.
We also made measurements of the number of photoelectrons detected, and rate capability
of the counters which are of general interest to the program.
 
In Section~2 we describe two  methods measuring $\sigma_{PMT}$.
Applications of these methods to conventional PMTs are described in Section~3. Results
for MCP PM 85001 from Burle are presented in Section~4.

\section{Methods for the measurements of PMT resolution.}

\begin{figure}
\begin{center}
\includegraphics[width=13cm,clip=true,bb=20 155 630 755]{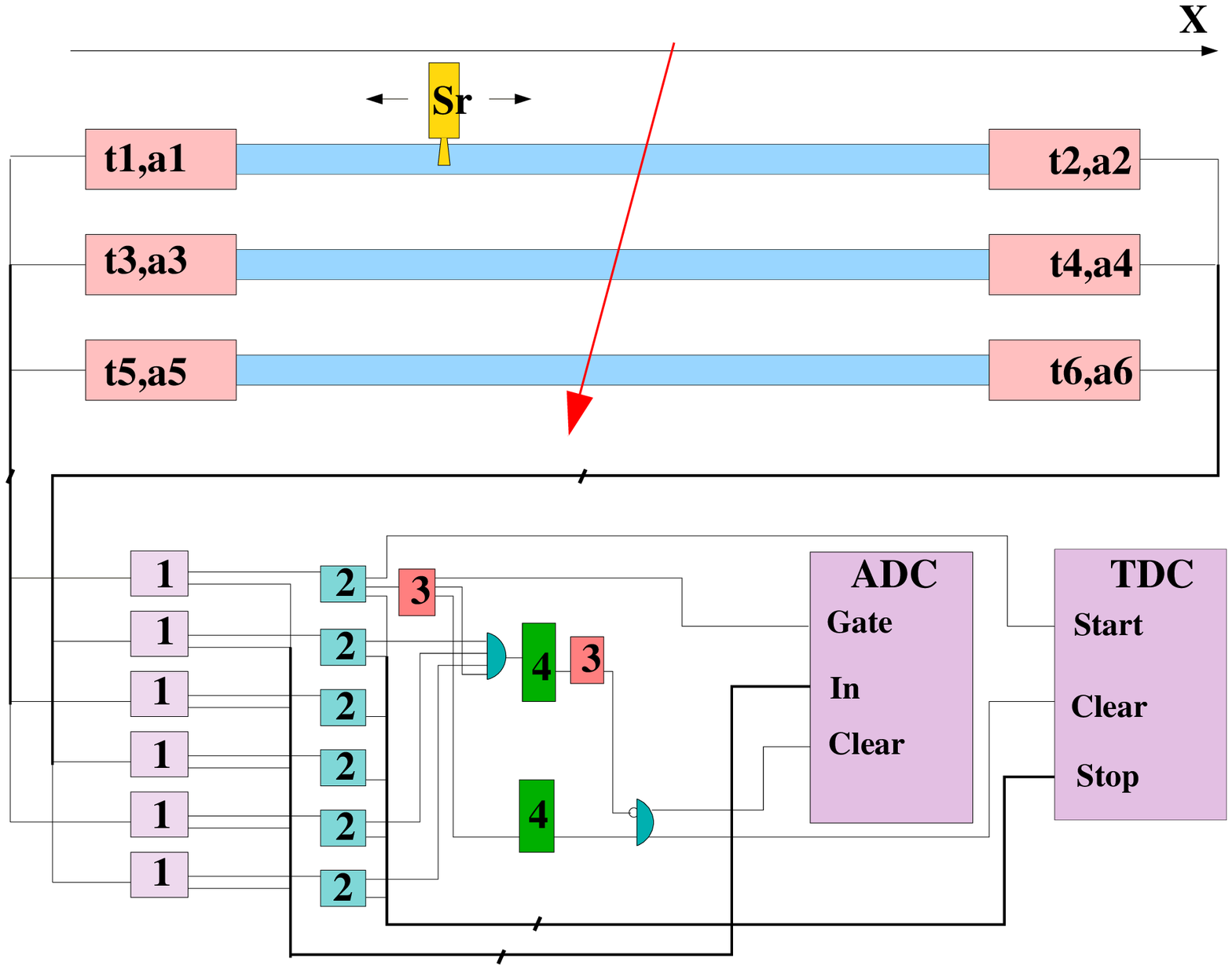}
\end{center}
\caption{Experimental  setup and electronics circuit diagram;
1-passive splitter, 
2-constant fraction discriminator, 3-fanout, 4-gate generator.} 
\label{camac}
\end{figure}

Historically, cosmic-rays have been a useful tool for 
studying the timing resolution of minimum ionizing tracks in test setups
\cite{giles}. We have used this procedure 
as a benchmark, and for checking the consistency
of a simpler and quicker method, ``coordinate method,'' detailed below.
First we review the technique for completeness. It requires at least
two identical counters, but we use a more robust method using three
stacked   parallel equidistant 
counters instrumented with 6 identical PMs.
The experimental setup and electronics circuit diagram are shown in Fig.~\ref{camac}.
To discriminate the PMT  signals  we use
constant fraction discriminators ORTEC-935. 
The signals were divided equally  at the inputs to the discriminators.
One part was used for discriminating at the threshold $\approx20~mV$,
while the second part was fed to ADC inputs.
The  arrival times $t_{1,...,6}$  of the  discriminated   signals relative to one of PMTs
were digitized by LeCroy-2228A TDCs.  The corresponding
pulses were integrated ($a_{1,...,6}$) within a $200~ns$ gate by LeCroy-2229 ADCs.

\paragraph*{The method of cosmic ray tracking.}

The times $t_{t,m,b}$ due to light flashes   
in  the top, middle and 
bottom- counters, respectively, are defined as: 
\begin{equation}
 t_t = (t_1+t_2)/2;~~~ t_m = (t_3+t_4)/2 ;~~~ t_b = (t_5+t_6)/2,
\label{tud} 
\end{equation}
where $t_{i=1-6}$ are the corresponding TDC readout values. 
The longitudinal coordinates  of the particle track $x_{t,m,b}$ 
we determine as
\begin{equation}
 x_t = (t_1-t_2)/2; ~~~  x_m = (t_3-t_4)/2;~~~ x_b = (t_5-t_6)/2.
\label{xud} 
\end{equation}
Thus the track may be reconstructed and straight trajectories may be selected.
For straight  tracks the  following relation between ideally measured $t_{t,m,b}$ 
holds: 
\begin{equation}
t_r=t_m - (t_t+t_b)/2 = 0.
\label{resid} 
\end{equation}
However, since  $t_{t,m,b}$  are smeared by  PMT resolution,
the value  $t_m - (t_t+t_b)/2$ ~jitters around zero, as well.
Hence, the  method is based on the statistical analysis 
 of   residuals of Eq.\ref{resid}:
\begin{equation}
\delta t_r = \delta \bigg((t_1+t_2)/4 -(t_3+t_4)/2 +(t_5+t_6)/4\bigg).
\label{resident} 
\end{equation}
Keeping in mind that one of the PMTs has been used for  the common
 start (~i.e.~the corresponding $\delta t=0$),
from  Eq.\ref{resident} we estimate the effective PMT resolution as
\begin{equation}
\sigma_{PMT} =\sqrt{var(t_i)}=\sqrt{\frac{16}{11}}\sqrt{var(t_r)}.
\label{eq2} 
\end{equation}

\paragraph*{Coordinate method.} 
In order to be able to perform measurements of the effective PMT resolution quickly,
we  have developed the  so-called  ``coordinate method.'' 
This method  requires only one scintillating counter  with two identical  
PMTs attached to its ends. 
The  resolution of the PMTs was determined from  the 
residuals of measured coordinates of ionizing 
$\beta$-particles from a radiative  $^{90}$Sr source. 
The  distribution of $\beta^-$  in the collimated beam  was derived from   
the counting rate profile \cite{r1} 
of the source.  The latter was measured 
with our scintillator by moving the source
in $1~mm$ steps across the scintillator and then  differentiated.
The resulting distribution was  fit with the Gaussian of 
 $\sigma_{beam}\approx3~mm$.
Such small  beam size may be  taken into account or even  neglected. 

The coordinate method is based on the simple  relation 
between the position $x$ of the source along the counter
and arrival times, $t_l$ and $t_r$, of signals at the two PMTs located  
at the ends of the scintillator:

\begin{equation}
 t_x = t_l-t_r =\frac{2x}{c_s}+const.
\label{eq:tx} 
\end{equation}
%\frac{1}{2}
where 
$x$ is the coordinate of the light flash, 
$t_x$ is the time interval measured with TDC,
$c_s\approx13.5cm/ns$ is the effective speed of light in scintillating media,
%$l$ is the length of scintillator;
the constant $const$ accounts for all kinds of propagation delays.
For convenience we define $X=x/c_s$ which measures position in  units of time.
%If the coordinates of csintillations are known, then the 

\begin{figure}
\begin{center}
\includegraphics[height=11cm,clip=true,bb=20 155 530 655]{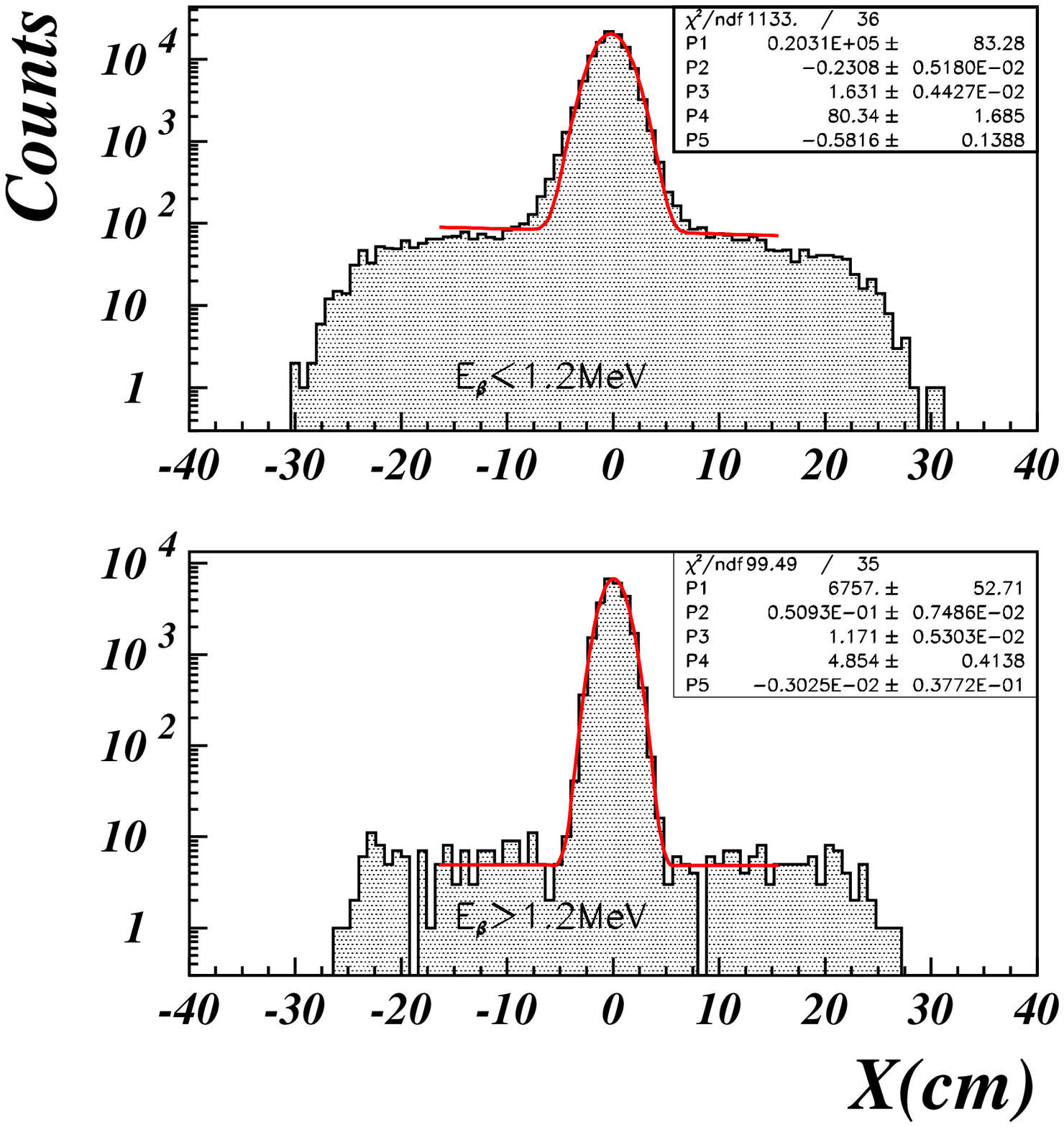}
\end{center}
\caption{ The ``$\beta$-ray''  images of $^{90}Sr$ source in two energy intervals of  $\beta$-particles.
Top panel:   $E_{\beta}<1.2~MeV$. 
Bottom panel: $E_{\beta}>1.2~MeV$. }
\label{xoffline}
\end{figure}

The variance of $t_x$ can be related to the PMT resolution $\sigma_{PMT}$ through the
relation
%(Eq.~$\ref{eq2}
%\begin{displaymath}
\begin{equation}
var(t_x)=var(t_l)+var(t_r)+var(t_{TDC})+\bigg(\frac{2}{c_s}\bigg)^2var(x)
\label{eq:vartx} 
\end{equation}
%\end{displaymath}
%
where 
$var(t_{l,r})$ are  the  variances for  left or right PMTs.
We assume that $\sigma_{PMT}^2$=$var(t_l)=var(t_r)$; the value
$var(t_{TDC}) \approx (25 ps)^2$ is the resolution of the  TDC measurements,
 and $var(x) \approx (15 ps)^2$ represents the size  of the irradiating  beam.
Thus, from this formula  one can  determine the single PMT resolution as 
\begin{equation}
\sigma_{PMT}=\frac{1}{\sqrt{2}}\sqrt{var(t_x)-var(t_{TDC})-
\bigg(\frac{2}{c_s}\bigg)^2 var(x)}   \approx   \frac{1}{\sqrt{2}}\sigma_{t_x},
\label{eq:spmt} 
\end{equation}
where the rightmost term  is obtained neglecting the beam size and  TDC resolution.
We illustrate the coordinate  method in Fig.~\ref{xoffline}, 
in which we show two images of the  radiative source. 
These images were accumulated in two energy intervals of $\beta^-$-particles, where
the longitudinal coordinates along the counter were determined via Eq.~(\ref{eq:tx}). 
The $50~cm$ wide  plateau  is 
the manifestation of  cosmic particles spanning by the counter. 
The two peaks at about zero are due to the ionization source 
placed at the center of the counter.

From the data in Fig.~\ref{xoffline} one can obtain a first estimate of the 
 resolution for minimum ionizing particles(MIP).
We determined the width
of the  peak  in  the bottom panel of this figure
 $\sigma_x=1.18~cm$.  From  Eqs.~\ref{eq:tx} and \ref{eq:spmt}  
we find $\sigma_{PMT}(1.2~MeV)$= $\sqrt 2 \times\sigma_x/c_s=123~ps$. 
The energy average deposited by MIPs is $4.4~MeV$ in our $2cm$ thick scintillator.
Therefore, since $\sigma_{PMT}\propto\frac{1}{\sqrt{E} }$, 
%the energy deposit of MIPs may be as high as $4.4~MeV$ in our scintillator, then
the expected 
 $\sigma_{PMT}(E_{MIP})\approx~\sqrt\frac{4.4}{1.2}\sigma_{PMT}(1.2)=64.2~ps$. 
We present this rough estimation to
demonstrate the transparency of the method.
However, for measuring  the  effective $\sigma_{PMT}$ we use the more rigorous coordinate method,
extrapolated to the energy of MIP, which we describe below.

\paragraph*{Coordinate method with  extrapolation  of $\sigma(E_{\beta})$ to $\sigma_{PMT}(E_{MIP})$.}

\begin{figure}
\begin{center}
\includegraphics[height=11cm,clip=true,bb=20 155 530 655]{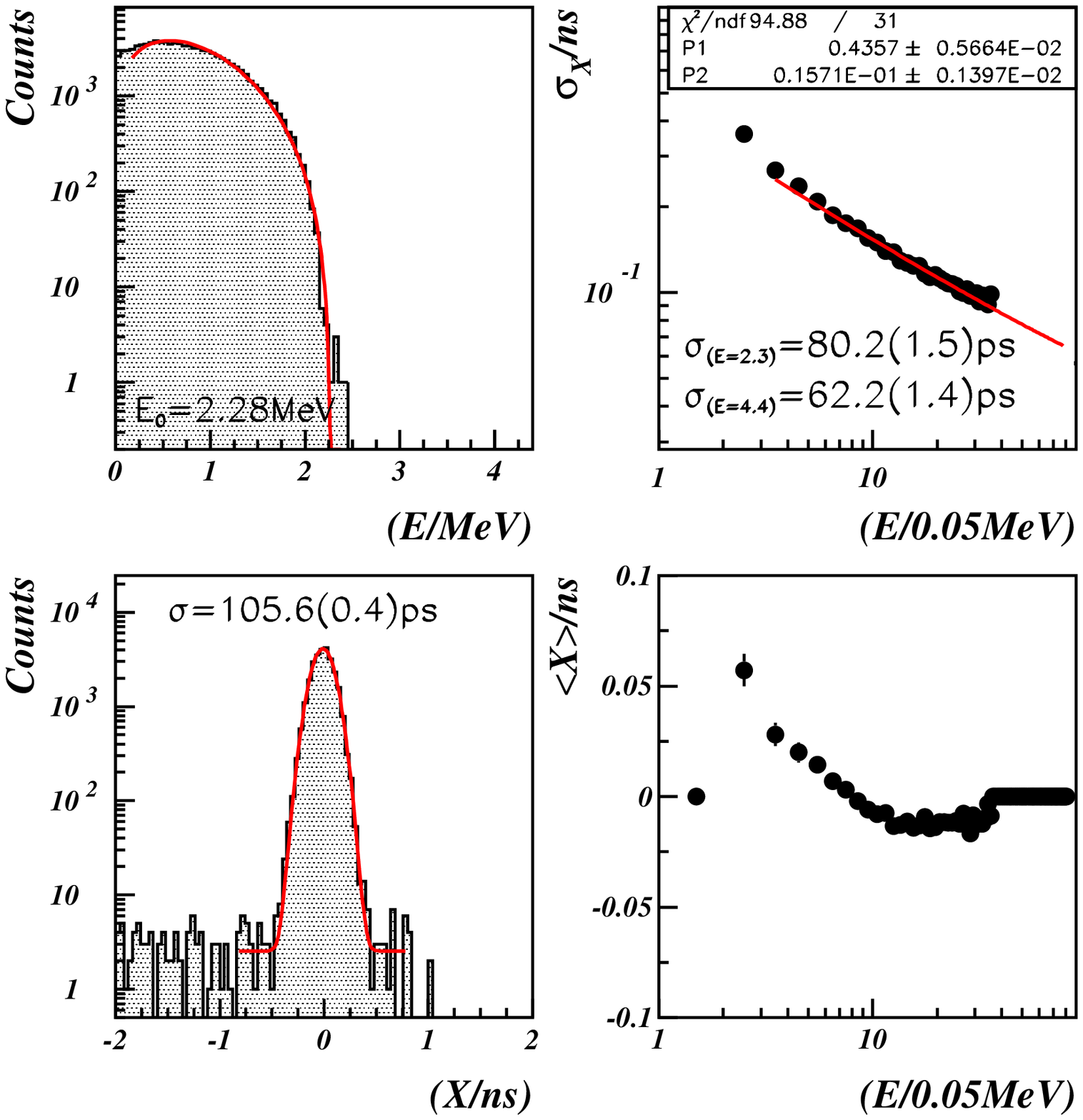}
\end{center}
\caption{Coordinate method for Hamamatsu $R2083$  at $^{90}Sr$  
location $-15~cm$.
Panel 
top-left: - ~energy ($E$) spectrum of $\beta$-particles ;
bottom-left:~coordinate($X$) spectrum of  $\beta$-particles; 
top-right:~$\sigma_X$ of the peak vs $\beta$-particle energy($E$);
bottom-right:~$\langle$$X$$\rangle$ of the peak vs $\beta$-particle energy($E$).
The peak position  on the $X$-scale was adjusted 
to zero for convenience.}
\label{srendep}
\end{figure}

From the off-line analysis of   data files accumulated  
with the  radiation source we determine  
the dependence of the PMT resolution upon the 
energy  of $\beta$-particles. The typical 
outcome  of such study with $^{90}Sr$ source is  
shown   in   Fig.~\ref{srendep}.
This figure
contains plots  for the  %one of two  
source location at $-15~cm$. %s: $+15~cm$ and $-15~cm$, respectively. 
Below we explain the contents of each panel.

The energy($E$) spectra (Fig.~\ref{srendep} top-left) of $\beta$-particles 
from the  $^{90}Sr$ source was determined using 
%(Eq.~$\ref{eq22}
\begin{equation}
E = k\sqrt{(a_l-p_l)(a_{r}-p_{r})}
\label{eq22} 
\end{equation}
where $a_{l,r}$ and $p_{l,r}$ 
are the amplitudes and pedestals from $ADC$s 
corresponding to left or right sides of the counter;  
$k$ is the calibration constant. The latter has  been determined from the fit of 
$\beta$-spectrum parameterized by  
%
%(Eq.~$\ref{eq3})
\begin{equation}
n(\varepsilon)=G(\varepsilon)\times
\varepsilon\sqrt{(\varepsilon^2-1)}(\varepsilon_0-\varepsilon)^2
\label{eq3} 
\end{equation}
where $\varepsilon=E/m_ec^2$,
$n(\varepsilon)$ is the number of events with specified $\varepsilon$, 
$\varepsilon_0$ = $2.28~MeV/m_ec^2$ is the upper limit of  $\beta$-spectrum for $^{90}Sr$,
$G(\varepsilon)$ is an emperical smoothing function.

The measured distribution of the longitudinal coordinate $X$, 
defined  by  Eq.~\ref{eq:tx}, is shown in Fig.~\ref{srendep} bottom-left.
The peak at zero is the image of the source.
Such distribution
can be plotted for a narrow slice in measured energy ($100~keV$ wide), as well.
In order to determine $\sigma_{X}(E)$,
we create a scatter  plot $X$~vs~$E$ via aforementioned slices.
The energy dependence of the centroid 
of the  source image $\langle$$X$$\rangle$($E$)
is determined by the above procedure and plotted in Fig.~\ref{srendep} bottom-right.
The width of the position distribution in each energy slice gives the measured
dependence of  $\sigma_{X}(E)$ and is plotted in Fig.~\ref{srendep} top-right.
As one can see from the plot, the data for $E < 2.3 MeV$ is well-described
by a $1/\sqrt{E}$ behavior, apparently  due to    
the increase of  the  number of photons. 
Extrapolating $\sigma_{X}(E)$ to  the energy deposit of 
minimum ionizing particles 
$4.4~MeV$ we estimate  $\sigma_{MIP}=63.9\pm1.5~ps$.

\paragraph*{Measurements of the number of primary photo-electrons.}

\begin{figure}
\begin{center}
\includegraphics[height=10cm,clip=true,bb=55 175 530 655]{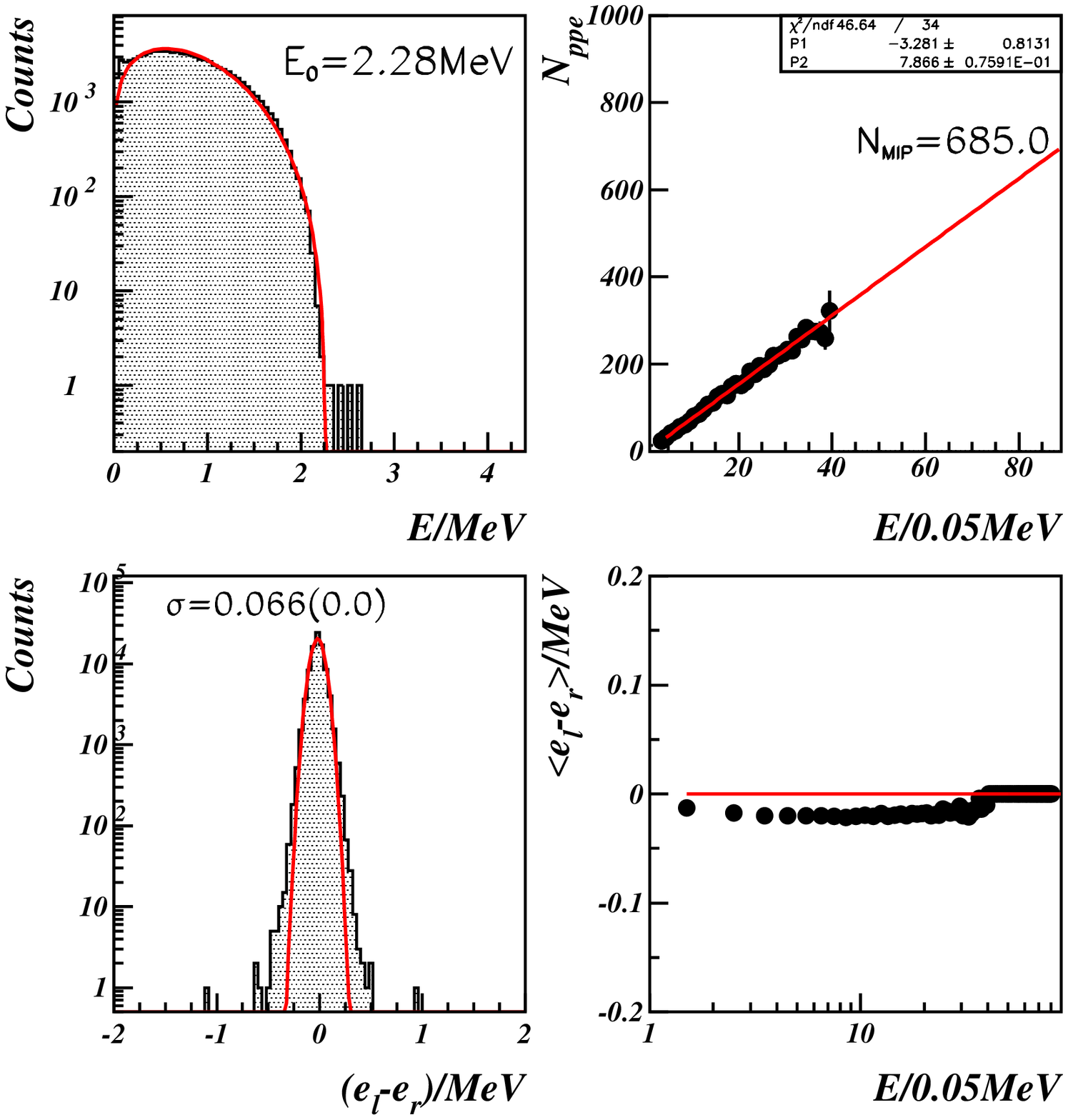}
\end{center}
\caption{
Top-left:~energy ($E$) spectrum of $\beta$-particles;
%\newline
top-right:~energy dependence of the number of primary photoelectrons 
$N_{ppe}(E)$. It is s expected to be $685\pm10$ at MIP 
energy of $\approx4.4 ~MeV$;
%\newline
bottom-left:~spectrum of $(e_l-e_r)$, where $e_{l,r}$ are 
the energies measured from two sides of the counter;
%\newline
bottom-right:~mean value of $(e_l-e_r)$ vs energy ($E$) of $\beta$-particles.}
\label{edopen}
\end{figure}

The following method has been used for estimating the number of primary  photo-electrons.
We define two energies \textit{measured} from  two sides of the counter: 
$e_{l,r}(E)=k_{l,r}\times(a_{l,r}-p_{l,r})$, where $ k_{l,r},a_{l,r},p_{l,r}$
 are the corresponding
calibrating factors,  ADC values and pedestals, respectively; obviously   $e_{l,r}(E)\propto E$.
% respectively $E=E_l-E_r$. 
%Both values are functions of the original energy deposit in the scintillator $E$.
We determine the total %\textit{measured} 
energy  $e$ and the difference  $\Delta e$ as 
%(Eq.~$\ref{elr})
\begin{equation}
e(E)=(e_l+e_r) \propto E~~~and~~~ \Delta e(E)=e_l-e_r.
\label{toten} 
\end{equation}
For the standard deviations of both values, defined above,  we write
%
%(Eq.~$\ref{elr})
\begin{equation}
%n(\varepsilon)=G(\sigma,\mu)
%(\varepsilon)\sqrt{(\varepsilon^2-1)}(\varepsilon_0-\varepsilon)^2
\sigma_{\Delta e}^2=
%\langle(\delta {e})^2\rangle=
%\langle(\delta e_1)^2\rangle+\langle(\delta e_2)^2\rangle=\sigma_e^2 .
var(e_l)+var(e_r)=\sigma_e^2 .
\label{elr} 
\end{equation}
Since  $\sigma_e^2\propto N(E)$, where  $N(E)$ is  the fluctuating number of primary 
photo-electrons in $both$ PMTs, we find
%
%(Eq.~$\ref{elr1})
\begin{equation}
%n(\varepsilon)=G(\sigma,\mu)
%(\varepsilon)\sqrt{(\varepsilon^2-1)}(\varepsilon_0-\varepsilon)^2
\frac{\sigma_e}{e}=\frac{\sqrt{N(E)}}{N(E)}.
\label{elr1} 
\end{equation}
Thus  the energy dependent number of primary photoelectrons may be determined as
%
%(Eq.~$\ref{elr2})
\begin{equation}
%n(\varepsilon)=G(\sigma,\mu)
%(\varepsilon)\sqrt{(\varepsilon^2-1)}(\varepsilon_0-\varepsilon)^2
%\frac{\sigma_e}{e}=\frac{\sqrt{N(e)}}{N(e)},
N(E)=\bigg(\frac{e(E)}{\sigma_e(E)}\bigg)^2=
\bigg(\frac{e(E)}{\sigma_{\Delta e}(E)}\bigg)^2
\label{elr2} 
\end{equation}
A special procedure is required to measure  $\sigma_e$ with the continuous energy spectrum. 
However, the function $\sigma_{\Delta e}(E)$ may be derived from the slices of the scatter plot 
$\Delta e$~vs~$E$. 
Using the fact that 
$\sigma_{\Delta e}$=$\sigma_e$ (Eq.~\ref{elr})
we  estimate  $N(E)$ via  Eq.~\ref{elr2} and extrapolate this function to MIP's energy. 
Thus measured  $N(E)$=685$\pm$10  and its extrapolation to the MIP energy of 4.4 MeV
is shown in Fig.~\ref{edopen}.
This number of primary photo-electrons collected in our  
$2\times3.3\times50~cm^3$ 
scintillator  is comparable to the number of $1000\pm100$ 
determined for  TOF counters  of CLAS\cite{r2} with a 
size of $5.1\times15\times32.3~cm^3$.
In our case, the number of primary photo-electrons is supposed to be  
emitted from \textit{two} photo-cathodes, the  sensitivity of which in $R2083$ is 
$\approx80~\mu$A/lm. 
The value   $1000\pm100$ relates to \textit{one} photo-cathode
of $EMI-9954B$ photo-multiplier with the 
sensitivity of $\approx110~\mu$A/lm. 

%If scaled to the thickness of our scintillator ($\times0.4$)
%and cathode sensitivity ($\times0.727$), the  CLAS number for $two(\times 2)$ PMTs turns to 
% $582\pm58$,
% for  \textit{two}  photo-cathodes of $EMI-9954B05$,
%which matches well to our result within  manufacturers tolerance.

\paragraph*{Advantages of the coordinate method.}

The coordinate method has several important advantages.
Firstly, it requires only one 
counter instrumented with two identical PMTs.
Secondly, the data taking and data analysis up to the final 
value takes only several minutes.
Finally, this method is insensitive to  the systematics related to the  
coordinate dependence of signal's timing, since the ionization is localized in 
the known  narrow region of $\pm0.3~cm$.
Moreover, the coordinate method  may be  used for 
studies of the coordinate 
dependent systematics, which  may be   measured directly  at different 
locations of the ionizing source.

\section{Measurements of time resolution.}
To discriminate the PMT  signals  we use the
constant fraction discriminators ORTEC-935. 
The signals were divided equally  at the inputs of discriminators.
One part was used for discriminating at the threshold $\approx20~mV$,
while the second part was fed to ADC inputs.
The  arrival times   of the  discriminated   signals relative to one of PMTs
were digitized by the LeCroy-2228A TDC.  The corresponding
pulses were integrated within $200~ns$ gate by the LeCroy-2229 ADC.

\begin{figure}
\begin{center}
\includegraphics[height=8cm,clip=true,bb=20 155 530 655]{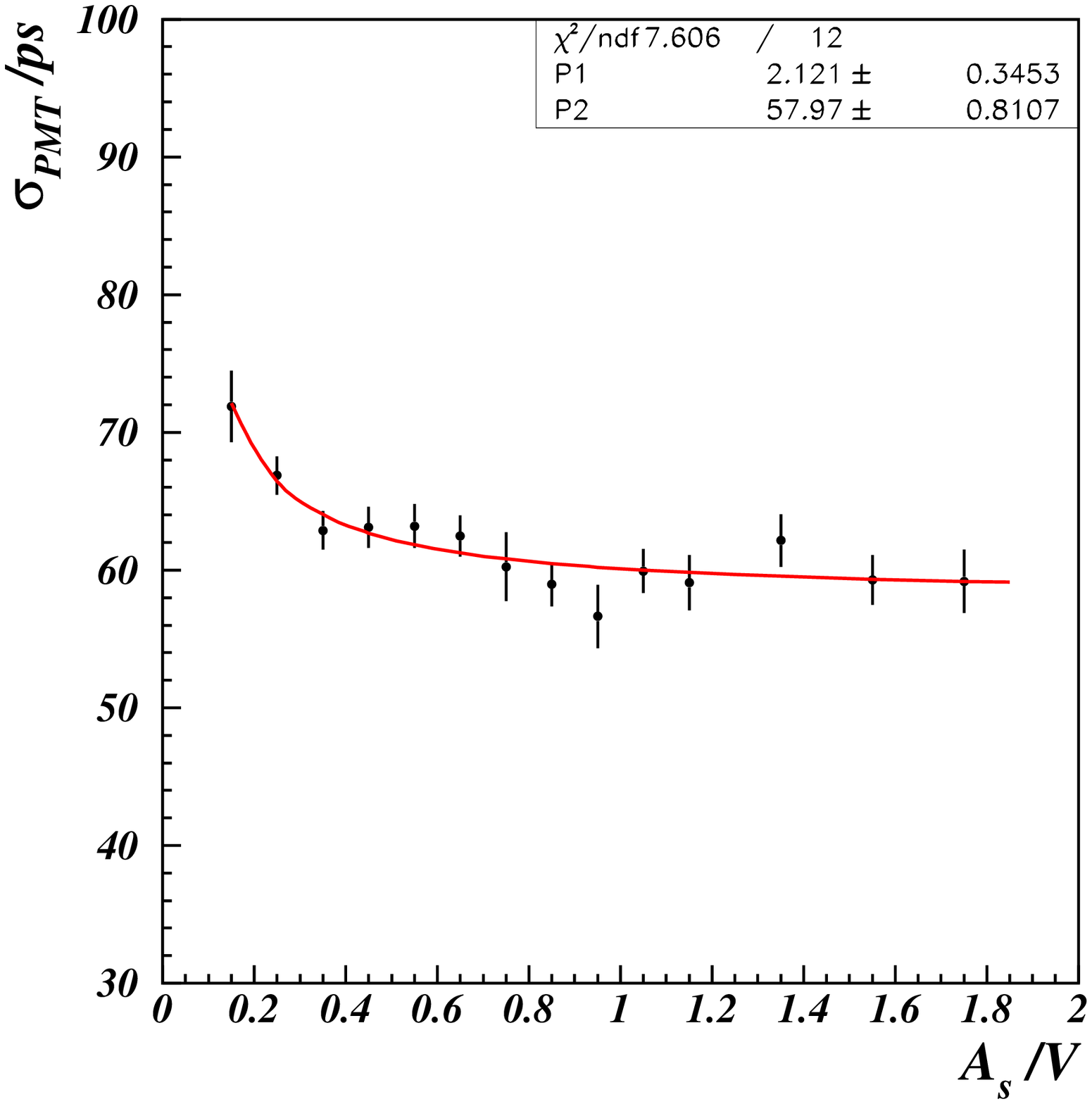}
\end{center}
\caption{ 
The dependence of $\sigma_{PMT}$ yielded by the  coordinate method  upon 
the signal amplitude $A_s$, which was measured with the  scope.
The mean value of $\sigma_{PMT}$ in the plateau region $(0.9,1.8~V)$ is of $59.5\pm0.7~ps$.
The curve represents the fit by $P1/A_s+P2$.}
\label{hvdep}
\end{figure}

\paragraph*{Study of $\sigma_{R2083}$ vs PMT gain via the coordinate  method.}
We have measured the dependence of 
$\sigma_{PMT}$ on the   PMT gain, which is obviously  proportional to the average amplitude of 
output signals. The HVs of both PMTs were tuned
to provide  amplitudes from both sides to be equal. 
The dependence of the resolution upon the 
amplitudes of signals is shown in Fig.~\ref{hvdep}. One can see that the resolution 
gradually improves up to 
the averaged signal amplitude of $900~mV$. 
Above  this value the resolution is 
almost constant.

\begin{figure}
\begin{center}
\includegraphics[height=11cm,clip=true,bb=20 155 530 655]{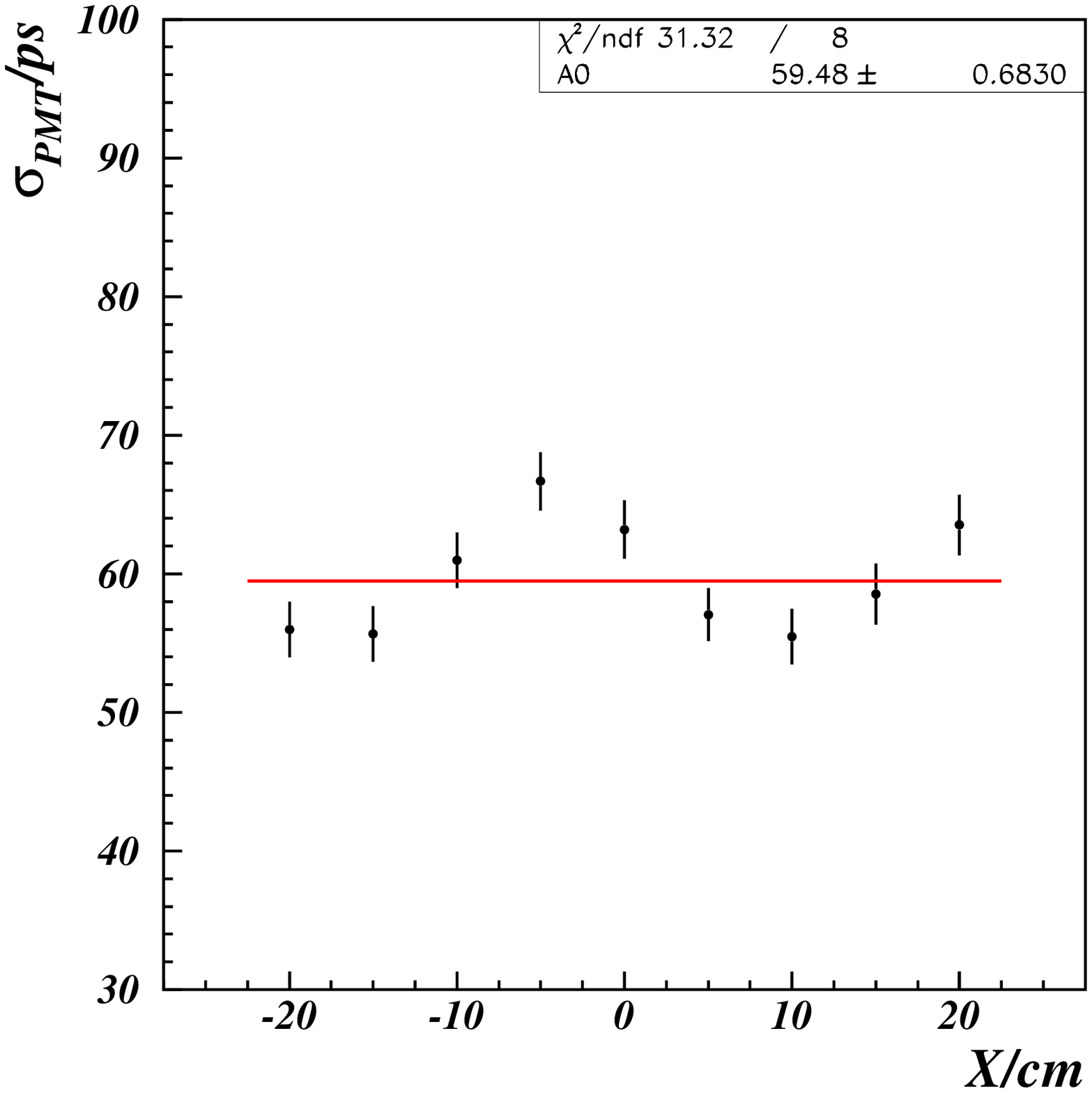}
\end{center}
\caption{ 
$\sigma_{PMT}$ vs  
$x$-coordinate at  voltages $2.37$ and $2.5~kV$ on the left and right PM, respectively. 
The mean  value is $59.5\pm0.7~ps$.}
\label{xdep}
\end{figure}

\paragraph*{Study of $\sigma_{PMT}$ vs position along the counter.}
In order to study the systematics of PMT timing
we have measured the $X$-dependence of both  the  peak location
and  $\sigma_{PMT}$. In advance
we have  equalized signals from both sides at the center of the counter.
The dependence of the resolution upon $x$-coordinate of the  source
is shown in Fig.~\ref{xdep}. 
The measured resolution oscillates near the mean value of $59.5\pm0.7~ps$.

\paragraph*{Cosmic-ray tracking and comparison of coordinate method.}

\begin{figure}
\begin{center}
\includegraphics[height=11cm,clip=true,bb=20 155 530 655]{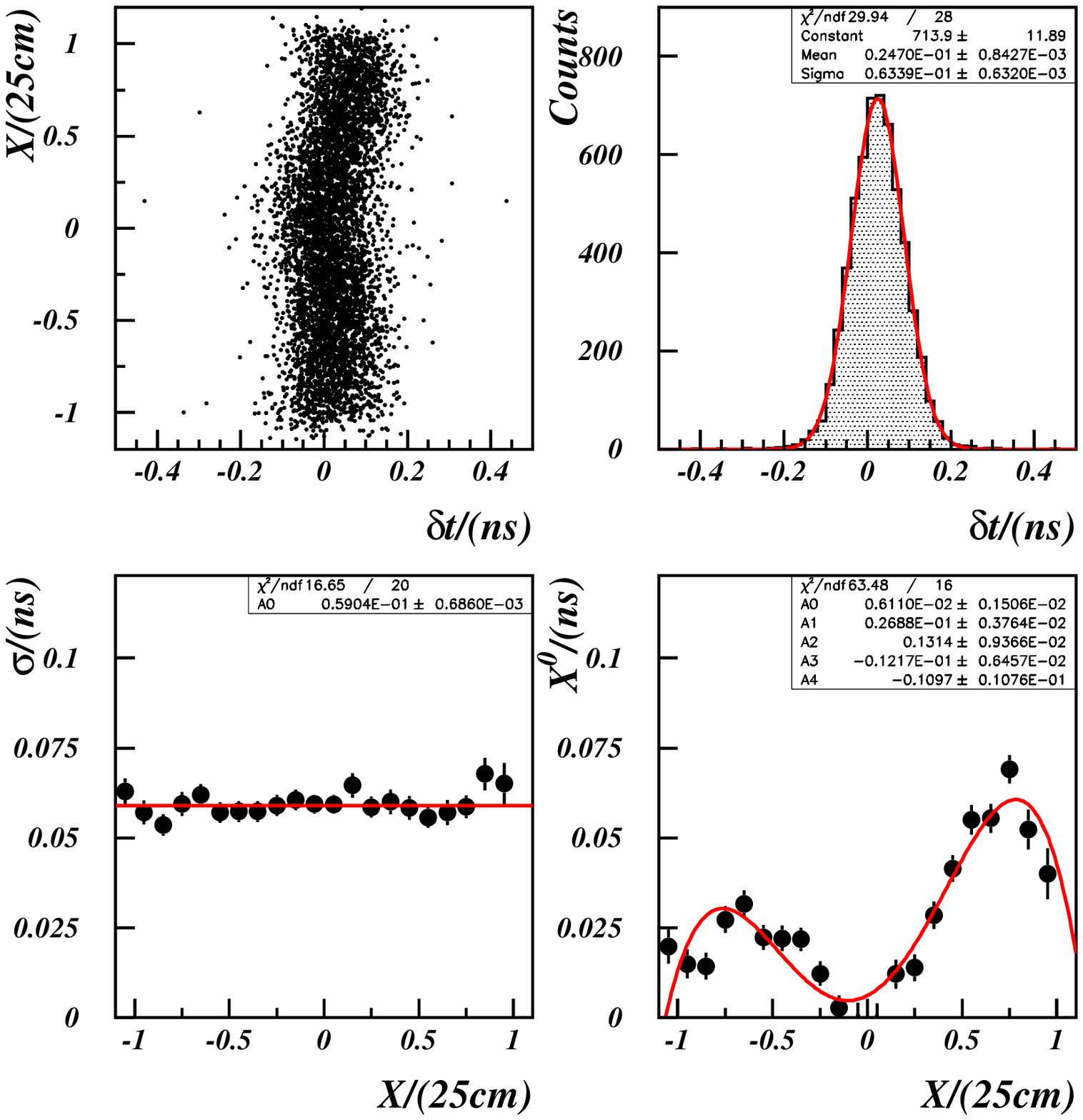}
\end{center}
\caption{
 Effective  PMT resolution  from the three counter method.
 %\newline
 Top-left:~scatter plot of 6 PMT residuals $\delta t $ vs $X$;
 %\newline
 top-right:~distribution of 6 PMT residuals $(\delta t)$, which yields the  overall $\sigma_{PMT}=63.4\pm0.6~ps$;
 %\newline
 bottom-left:~second moment of residuals $\sigma$ vs $X$, which yields the  local 
 $\sigma_{PMT}=59.1\pm0.7~ps$ as mean value;
 %\newline
 bottom-right:~First moment of $\delta t$ distribution $X^0$ vs $X$. 
 %\newline
         }
 \label{cosm0504}
 \end{figure}

The result from the three counter tracking  method is shown in Fig.~\ref{cosm0504}.
The local resolution  has been determined to be of 
$59.1\pm0.7~ps$. The overall resolution yielded by this method is of $63.4\pm0.6~ps$.
The last value  is  worse due to  some $x$-dependent displacement  
of residuals, which may be  seen in Fig.~\ref{cosm0504} bottom-right.

Thus, we consider that the coordinate method is well established. It yields values which
agree nicely with
the tracking method applied to  cosmic particles. 
Therefore, this quick method may be trusted for studies of
timing resolution of various photo-multipliers. 

\section{Resolution, counting rate, and number of photo-electrons for MCP MPs.}
We have implemented the coordinate method to measure the resolution of Micro-channel 
Plate PMs ``Burle-85001-501'' with the 
same scintillator and setup.
%using the same scintillator and  methods.
%It looks like  the effective number of primary electrons participating in the 
%avalanche development is of about factor $5.7^{-1}$ lower  
%against the PMs of Hamamatsu. 
%Therefore the timing resolution of MCP PM is approximately $\sqrt{5.7}=2.39$
% times worse.
However, an amplifier ($LeCroy-612A$) was required for these measurements, 
since signals from the MCPs are too small for   discriminating.
%The setup is shown in Fig.~\ref{mcpsetup}
We note that the sensitive surface of the
``Burle-85001-501'' assembly  is formed by  four MCPs.
First we performed the resolution tests with four MCPs connected to the input circuit
in parallel. 
%In this case   the effective square of the cathode is almost 
%equal to the size of our scintillator.
Then we repeated  our tests with only one of four MCPs feeding the input circuit.
We find no difference between two measurements. Thus the individual transition 
times and deviations may be considered  equal. 
We emphasize that according to its data sheet the photo-cathode
sensitivity of $Burle-85001-501$ 
lies between  $40$ and 55$~\mu$A/lm.
% which is twice lower thah that of $R2083$
This value is about twice  lower than that for $R2083$ PM. 
%of Hamamatsu.
 Therefore the number of primary photoelectrons has 
to be at least twice lower and one can expect  
the timing resolution of MCP PM  to be  of  $\sqrt{2}$ times worse.
%
%
%\begin{figure}
%\begin{center}
%\includegraphics[width=.60\textwidth]{mcpsetupview1.eps}
%\caption{Photograph of the MCP setup.}
%\label{mcpsetup}
%\end{center}
%\end{figure}
%

\begin{figure}
\begin{center}
\includegraphics[height=8cm,clip=true,bb=20 155 530 655]{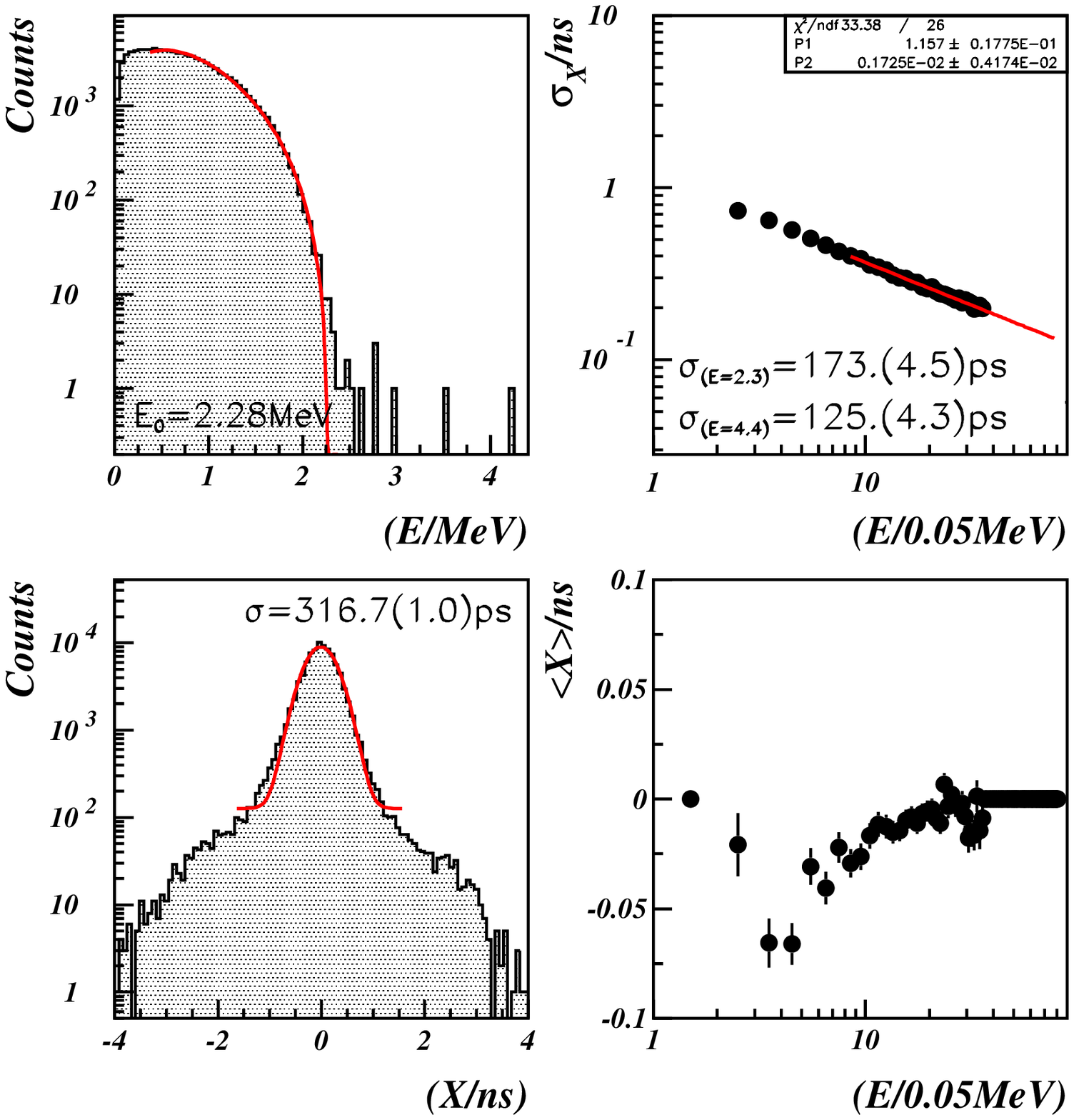}
\end{center}
\caption{Coordinate method with MCP PM from ``Burle'' at highest  $HVs=(2150~V,2400~V)$.
Extrapolated to MIPs $\sigma_{PMT}$ is of $125\pm4~ps$.
Amplification factor is of $10^1$.}
\label{x+mcp}
\end{figure}

The resulting plots of  coordinate method with MCP~PM are shown  in Fig.~\ref{x+mcp}. 
The method yields  $\sigma_{PMT}=125\pm4$ for MIPs. This value is   
almost twice  the resolution  
of the $R2083$. Since the single electron  resolution of MCP is very good then
the reasons for the worse resolution could be: (1) noise of preamplifier,
(2) increased effects of  statistical fluctuations of primary photo-electrons.
The first assumption  will be checked  in future measurements with an  ``on-board'' preamplifier, 
such as that described in Ref.~\cite{popov}.
% It has to be located as
% close as possible to the MCP output (on-board PM such as 
% described in Ref.~\cite{popov}). 
We checked the  second  possibility by measuring  the number of 
primary photoelectrons.

\begin{figure}
\begin{center}
\includegraphics[height=8cm,clip=true,bb=20 155 530 655]{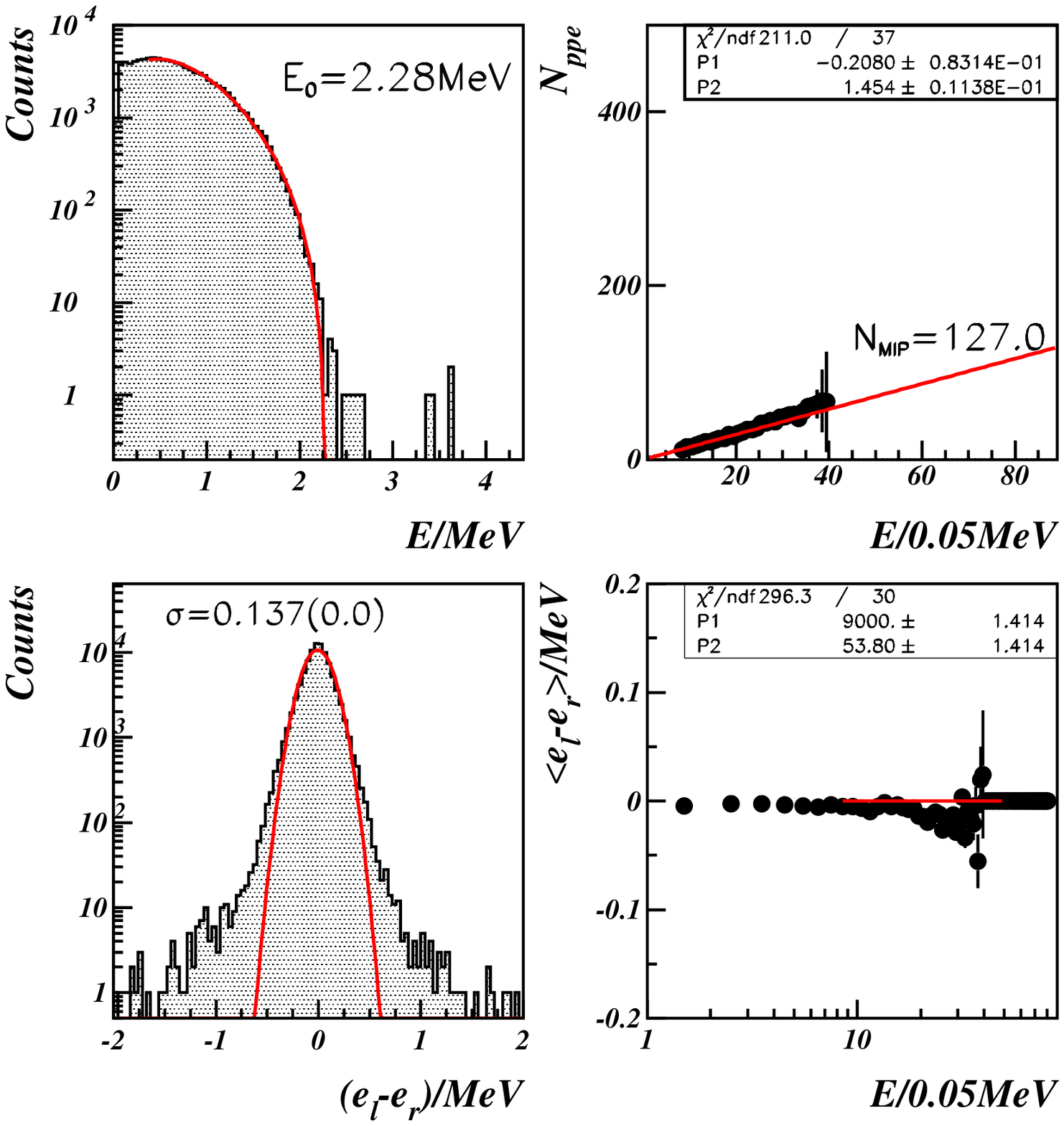}
\end{center}
\caption{Coordinate method with MCP PM from ``Burle'' at $HVs=(1940,2100)~V$.
Extrapolated  to MIPs  $N_{ppe}$=$127\pm10$.
Amplification factor is of $10^2$.}
\label{mcpnppe}
\end{figure}

\paragraph*{Number of primary photoelectrons in the MCP.}
We have measured the number of 
primary photoelectrons produced by M.I.Ps
%in the same manner as we 
using the same method as
%have done that 
for the PMs from Hamamatsu. 
The resulting distributions 
are shown 
in Fig.~\ref{mcpnppe}.
The  extrapolated  $N_{ppe}(4.4~MeV)$ (for $two$ MCPs)  was found to be of  $127\pm10$.
This value is $5.4$ times lower compared to the $R2083$ photomultiplier
and about $2.7$ times lower than expected from the data sheet value. 
We note that  $\frac{\sigma_{85001}}{\sigma_{R2083}}\approx 2.2$, which 
 agrees with the  ratio  of corresponding $\sqrt N_{ppe}$ numbers$(2.3)$. 
Thus, our measurements look consistent.

\begin{figure}
\begin{center}
\includegraphics[height=8cm,clip=true,bb=20 155 530 655]{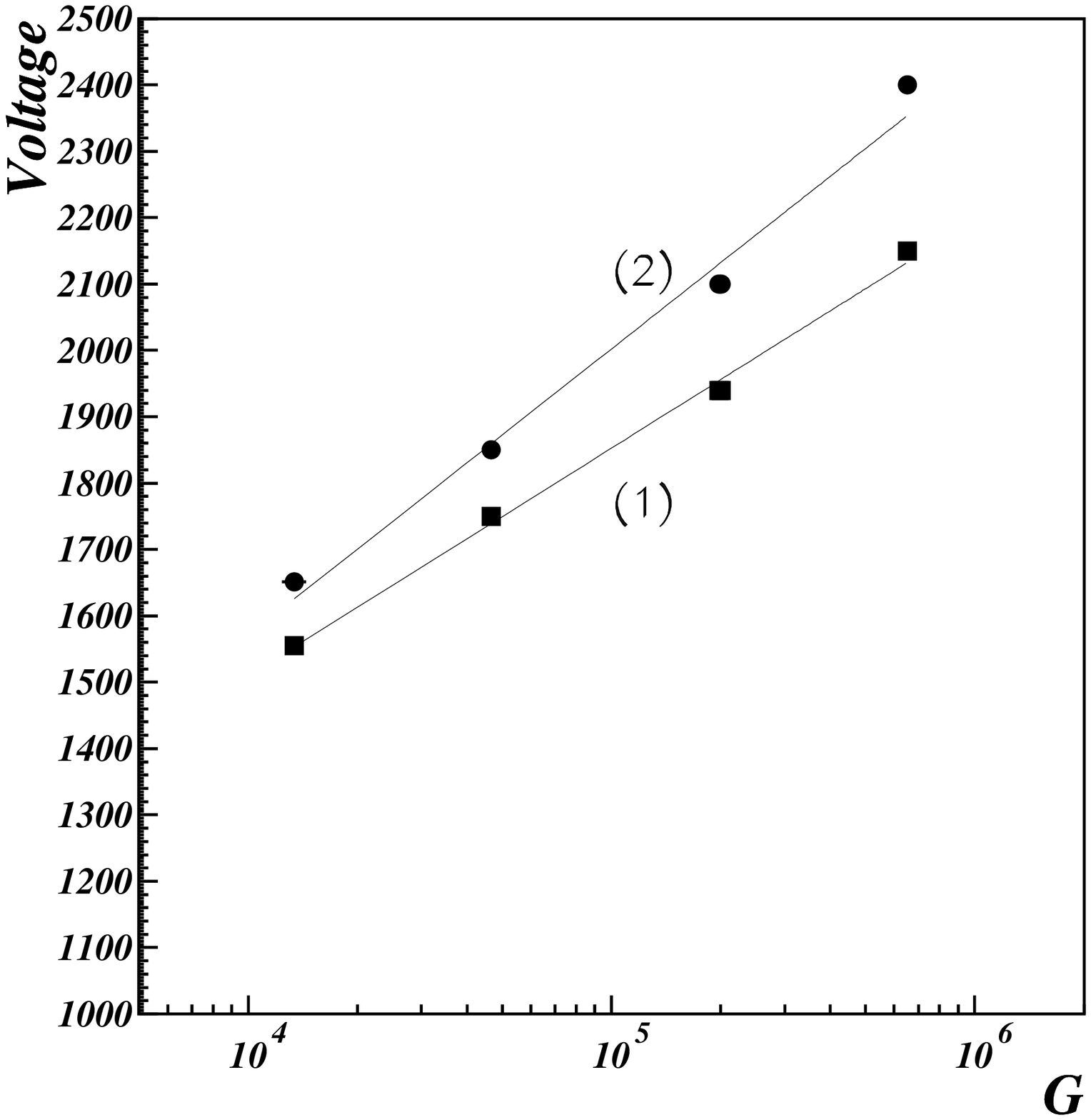}
\end{center}
\caption{High Voltage  vs MCP Gain.
The curves (1) and (2) are fits to the power law.
(1)-left MCP PM,(2)-right MCP PM.}
\label{hvvsgain}
\end{figure}

\paragraph*{Measurements of MCP gain.}
The MCP gain  was determined using the measured  $N_{ppe}$ and the charge 
spectra of $\beta$-particles measured by ADCs with the source at the center of the counter. 
We can thus plot the  measured gains as function of the MCP voltages, as shown in Fig.~\ref{hvvsgain}.
We remind the reader that the voltages have been tuned to provide equal
amplitudes from two sides of the detector.

\paragraph*{Timing resolution vs MCP gain.}

\begin{figure}
\begin{center}
\includegraphics[height=8cm,clip=true,bb=33 175 530 680]{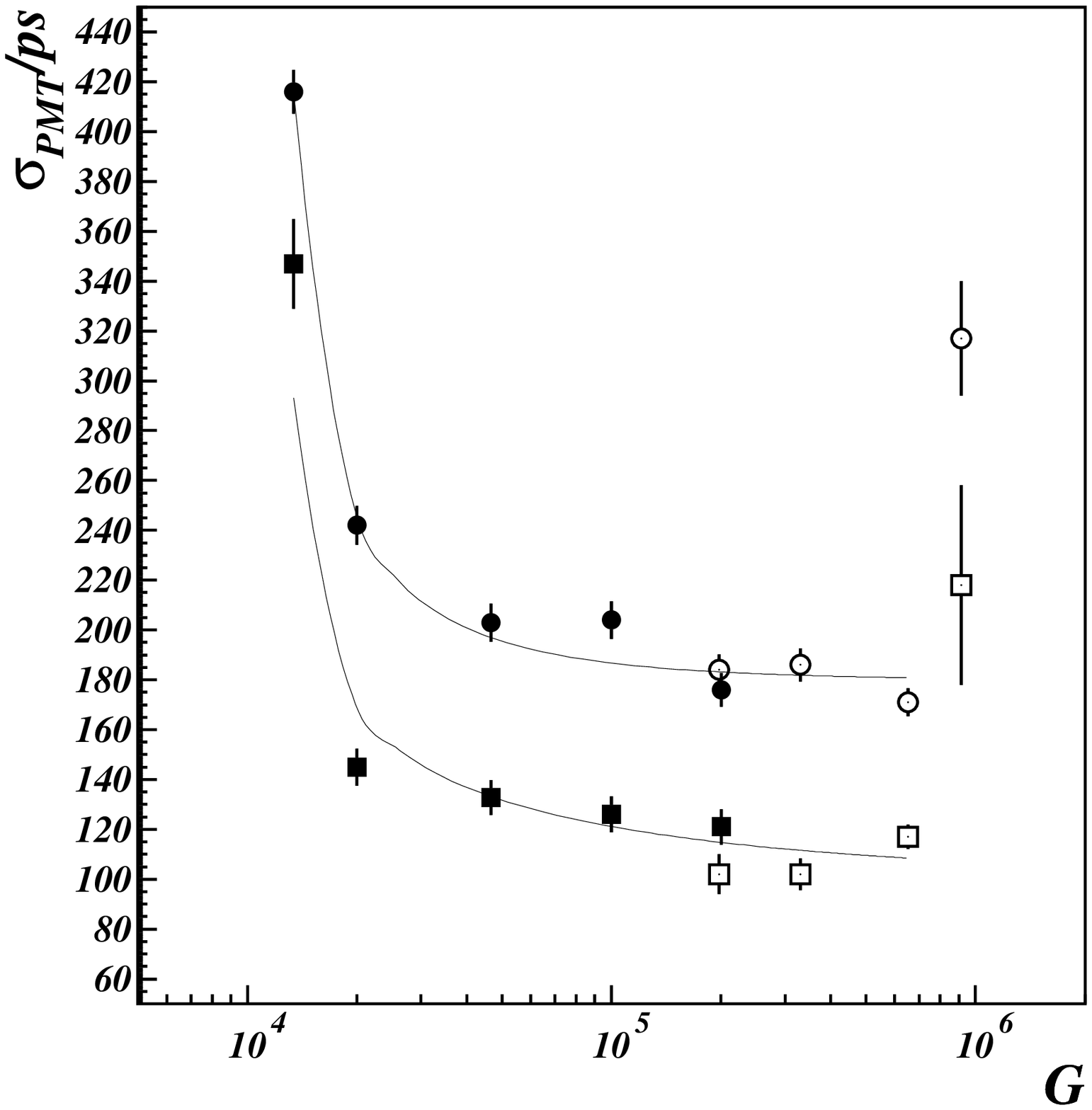}
\end{center}
\caption{Resolution  vs MCP gain(G). Squares are for $\sigma_{PMT}$ at $\Delta E=2.28~MeV$.
Circles are for $\sigma_{PMT}$ extrapolated to  $\Delta E=4.4~MeV$(MIPs)
The curves are  fits to the $G^{-2}$ dependence.  
Amplification factor varies from  $10$ in the gain region of$~(2.0,6.5)\times10^5$(white symbols) to $10^2$ in the 
region $(0.13,2.0)\times10^5$(black symbols).
Two  points at  $2.0\times10^5$ were measured  at amplification $10^1$ and $10^2$ for comparison.}
\label{sigmamip}
\end{figure}

We have measured the effective resolution of MCP PMs at different high voltage settings and
amplification factors $10^{0},10^{1},10^{2}$ cascading our preamplifier's.
The resolution yielded by 
coordinate method  is shown in Fig.~\ref{sigmamip} as a function of the MCP gain.
One can see a plateau   between gains $0.4\times10^5$~and~$6.5\times10^5$,
 where the extrapolated resolution lies  in the 
interval $(102,132)~ps$.
The leftmost and rightmost points in this figure were obtained at very low output signals (below
$5pC$) with the amplification factors  of  $10^2$ and $10^0$, respectively. Therefore we exclude them 
from  consideration.

\paragraph*{Counting rate capability of  MCP PM with Light Emitting Diode.}
The counting rate capability of  85001 PM  has been measured in the following way.
The radiative source was replaced  by a Light Emitting Diode.
This LED was fed from the pulser $LeCroy-9210$ with a pulse of $5V$ 
amplitude and controllable width $50<$$\tau$$<100$$~ns$. 
From the scope measurements the rise time of the
light signal was found to be equal to $\tau$, while  
%Therefore timing measurements with $X+$ method with LED makes no sense, since the 
%distribution is very wide(of about $20~ns$). 
%However, the signal produced by LED 
%fits well into the ADC gate($200~ns$) and reliable ADC measurements of the pulse charge are possible. 
%According to the measurements with a scope 
the amount  of light  produced by LED $\propto \tau^2$.

\begin{figure}
\begin{center}
\includegraphics[width=13cm,bb=35 175 528 678,clip=true]{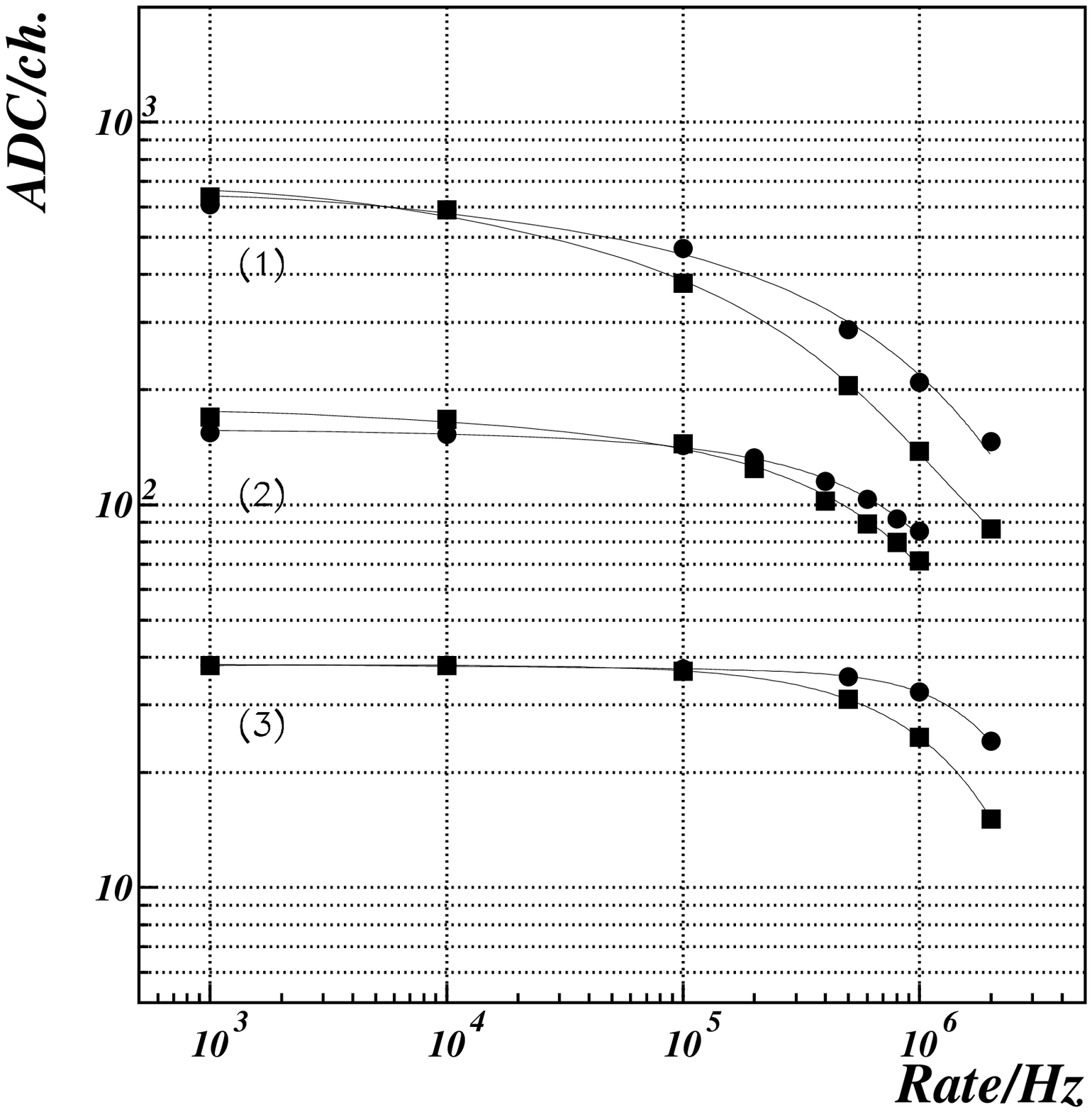}
\end{center}
\caption{The integrated current of MCP~PM signal (ADC counts) vs 
the rate of light flashes  at: 
%
%\newline
(1) HVs=$(2400,2400)~V$ and $\tau=100~ns$;
%\newline
(2) HVs=($2400,2400)~V$ and $\tau=50~ns$;
%\newline
(3) HVs=$(1815,1875)~V$ and $\tau=100~ns$.
%\newline
Amplification factor is $10^2$~(two cascaded amplifiers).
The light has been generated by the LED fed with the pulser.
}
\label{crcpb}
\end{figure}

We tested the counting rate capability at MCP gains of approximately $6.5\times10^5$ and  $6.5\times10^4$.
%at $HVs=(2400~V,2400~V)$ and $(1815~V,1875~V)$ for the left and right PMs, respectively.
At high gain we have also  taken data at \textit{high} and \textit{low}  
intensities of light flashes, corresponding to  
$\tau$=$100~ns$ and $50~ns$,  respectively.
The behavior of  signal charge vs counting  rate is shown in Fig.~\ref{crcpb}.
From curves (1) and (2) in this figure one can conclude  that indeed the signal charge 
is proportional to $\tau^2$.  
The plateau region of curve~(3), obtained with  $low$ gain   
$\approx0.65\times10^5$, 
is significantly wider than that of curve~(1) at  $high$ gain  $\approx6.5\times10^5$.
%We note that at HVs=(1815,1875)~V the gain is of about 15 times lower than at HVs=(2400,2400)~V.
 From    Fig.~\ref{sigmamip}   we find that 
$\sigma_{85001}\approx130~ps$ at gain $\approx0.65\times10^5$.

\begin{figure}
\begin{center}
\includegraphics[height=11cm,clip=true,bb=20 155 530 655]{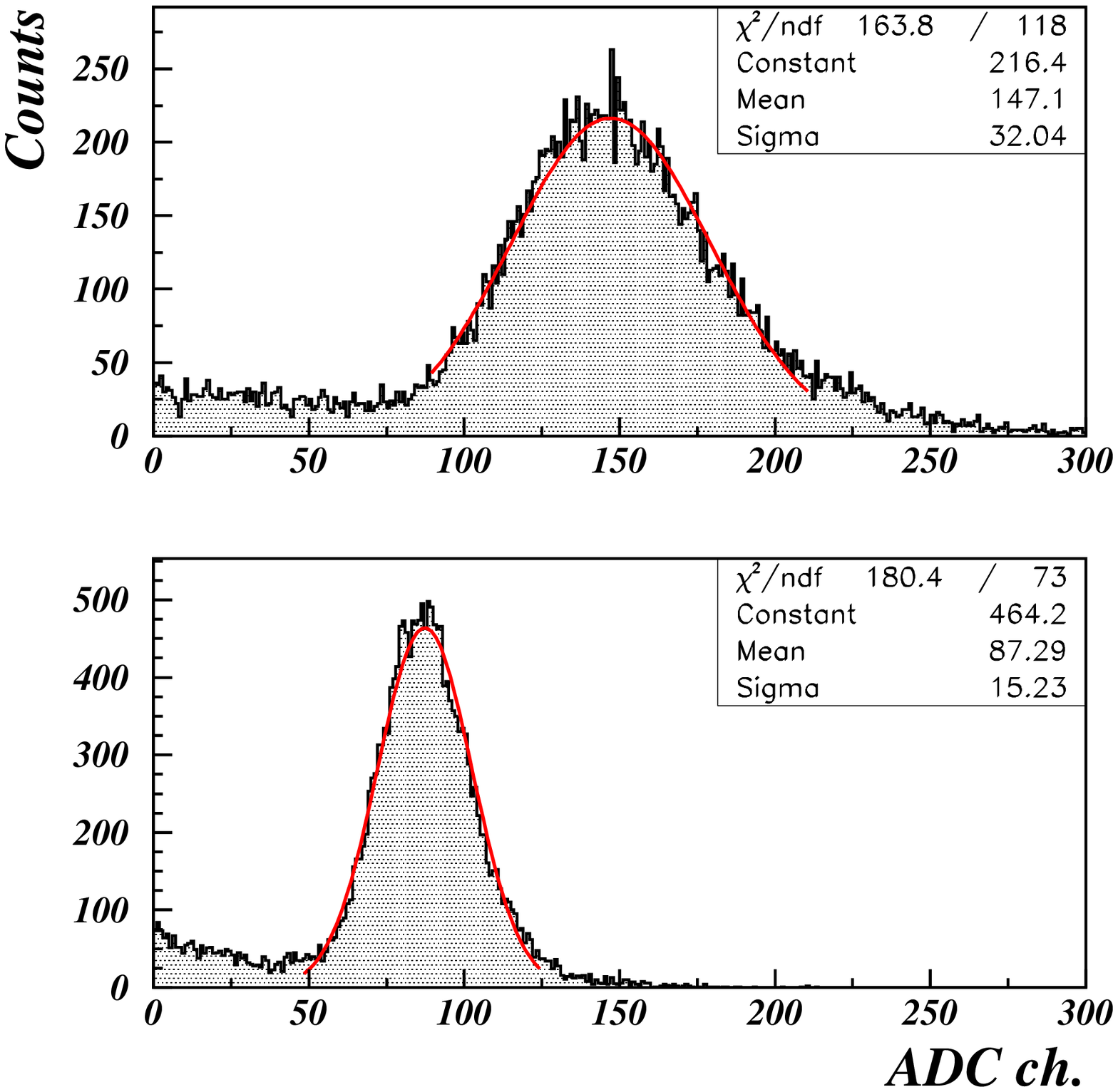}
\end{center}
\caption{The pedestal subtracted ADC spectra of signals from LED operating  at $2~MHz$ rate. 
The amplification to MCP signals  is  $10^2$, HV is $2.4~kV$ for both PMs.}
\label{mcpadcsp}
\end{figure}

Hence, one can conclude that at $low$ gain, 
with an amplification factor of $10^2$, the MCP PM  can operate with a resolution of $\approx130ps$
and at counting rates up to $0.5\times10^6~Hz$.
The last value corresponds to about $75\%$ of the gain at $10^4Hz$.   
To illustrate  the performance of 85001 MCP PM   we show  in  Fig.~\ref{mcpadcsp} 
two  ADC spectra   at $very$~$high$ counting rate of $2\times10^6~Hz$ and ultimate HV of $2400~V$.
Despite of severe conditions both spectra contain a clear peak corresponding to LED signal.

\paragraph*{Number of primary photoelectrons from LED.}
It is important to know the energy of the light flash at the input of the PM.
We have estimated the number of primary photoelectrons produced by the  LED 
in both MCP PMs.
For our measurements of $N_{ppe}$  we have used ADC spectra obtained
at  counting rate of $10^4$ and gain $\approx6.5\times 10^5$. 
From several measurements of the  peak width (as in Fig.~\ref{mcpadcsp})
we have estimated the total number of 
primary photo-electrons produced by LED as  $114\pm25$. This value is close to 
the $N_{MIP}$  from  Fig.~\ref{mcpnppe}. Therefore, we consider the results shown in 
Fig.~\ref{sigmamip} as estimations of the 85001 counting rate capability  for 
minimum ionizing particles.

\section{Conclusions and Discussion}
With two different methods we confirmed that the effective 
timing resolution of Hamamatsu $R2083$, attached directly to our  scintillator, is better than $60~ps$.
This value establishes the basis for further studies  with $1m$ long light guides, from which we
expect the effective resolution to be significantly worse.
Therefore, we  are looking for alternatives to such a  design.
With this purpose in mind
we have measured the  timing resolution  for the $Burle~85001-501$ MCP PMs. 
The value of resolution  $130\pm4~ps$ was obtained  at the very low  gain 
$\approx 4.5\times 10^4$ using an external  amplification
of $10^2$ to the  PM's signal. 
We compare these measurements in Table~\ref{table1}.
\begin{center}
\begin{tabular}{|l|c|c|c|r|} \hline
Method            & Acceptance  &Particles        & $\Delta$$E$      &$\sigma_{PMT}~~~$         \\ \hline
6  R2083      &             &                 &                  &                       \\
tracking        &$0\pm25~cm$  &  cosmic              &      $4.4~MeV$       & 63.4$\pm$0.6$~ps$          \\ \hline
6  R2083      &             &                 &                  &                       \\
tracking          & $<local>$   &  cosmic            &    $4.4~MeV$  & 59.1$\pm$0.7$~ps$        \\ \hline
$X$+,~R2083             &             & $\beta~from~^{90}Sr$                &$extrapolated   $ &                       \\
~        & $<local>$  &  $E_{\beta}<2.28~MeV$      &   $to~\approx4.4~MeV$   & 59.5$\pm$0.7$~ps$        \\ \hline
$X$+,~85001    &             & $\beta~from~ ^{90}Sr$                &$extrapolated   $ &                       \\
~MCP~PM  & $<local>$  &  $E_{\beta}<2.28~MeV$      &   $to~\approx4.4~MeV$   & 130$\pm$4$~ps$        \\ \hline
\end{tabular}                                                                      
\end{center}
\begin{table}[htbp]
\caption{PMT resolution(standard deviation) 
obtained using  different methods. $X+$ stands for the coordinate method extrapolated to MIP energy.
The errors shown in this table 
are obtained by   fitting procedures. We estimate a possible systematic error 
to be of about +$5\%$ to the shown values.}
\label{table1}
\end{table}
We emphasize that the results shown in Table~\ref{table1} were obtained with the
prototype, which has no   light guides. A good agreement between  
$\sigma_{R2083}$'s from the cosmic ray method 
($6~PMT$ tracking) and  coordinate method with extrapolation to higher energies  ensures 
that coordinate  method is an 
adequate tool for measuring of $\sigma_{PMT}$ at MIP energy.

Another important parameter of MCP PM is its counting rate capability, 
which  was also addressed in our studies.
We have shown that at low  MCP gain  the counting rate of  
MIPs can be as high as $0.5\times10^6$.

The resolution is dictated  mostly by fluctuations of primary electrons.
Within  coordinate method  we have estimated  the  
number of primary photo-electrons caused by MIPs in MCP. 
This number is 2.7 times lower than  the value expected from the $Burle-85001$ data sheet. 
However, the same method of estimating the number of primary photo-electrons
gives the  reasonable value for the $R2083$ PMs  of Hamamatsu.
Therefore, we believe that the resolution of the
$Burle-85001$ MCP~PM should improve by  $\approx 1.6$ to about $80~ps$
provided  the number of primary photoelectrons corresponds to the data sheet for the
85001 photo-cathode. 
In that case the timing resolution of $Burle-85001$ would  come  significantly closer to  
the requirements  of  CLAS experiments at $12~GeV$. Its current counting rate capability ($0.5\times10^6$) 
with external  amplification of  $10^2$
is also quite close to the typical counting rates of CLAS experiments.
Nevertheless, we believe that both the resolution and  counting rate capability of MCP PMs has to be  improved.
The progress  could be achieved   
via  implementation of  Wide Dynamic Range MCPs 
such as F6584 from Hamamatsu. Photo-multipliers instrumented with WDR~MCPs 
could  operate at significantly higher counting rates and/or gains, and resolution, as well. 
Therefore, we encourage  PMT manufacturers to develop such photo-multipliers.

%\newpage
%\input{appendix.tex}

\ack
%We would like to acknowledge the outstanding efforts of the staff of the 
%Accelerator and the Physics Divisions at JLab that have contributed to
%the design, construction, installation, and operation of the CLAS detector.
%This work was supported 
%in part by the Istituto Nazionale di Fisica Nucleare, the 
%French Centre National de la Recherche Scientifique, 
%the French Commissariat \`{a} l'Energie Atomique, the U.S. Department of Energy, the National 
%Science Foundation, and the 
%Korean Science and Engineering Foundation.
%The Southeastern Universities Research Association (SURA) operates the 
%Thomas Jefferson National Accelerator Facility for the United States 
%Department of Energy under contract DE-AC05-84ER40150.

The Southeastern Universities Research Association (SURA) operates the 
Thomas Jefferson National Accelerator Facility for the United States 
Department of Energy under contract DE-AC05-84ER40150.
This work was also supported in part by the 
Korea Research Foundation.
%% LaTeX2e file `CLAS_nim.bib'
%%
% bibliography for CLAS_nim

%\begin{figure}
%\begin{center}
%%\includegraphics[width=13.2cm,clip=true,bb=0 0 325 425]{class_pp_3pic.eps}
%\includegraphics[width=13.2cm,clip=true,bb=-300 -300 825 1025]{CLASTOF.ps.gz}
%\end{center}
%\caption{ Conceptual design of the central  time-zero counter for the 
% CLAS upgrade at 12~GeV.}
%\label{clas}
%\end{figure}
%%Fig.~\ref{clas}

%\begin{figure}
%\begin{center}
%\includegraphics{file=CENT_TOF_AUG.PS.gz,
%bbllx=0pt,bblly=180pt,bburx=595pt,bbury=600pt,width=.60\textwidth}
%\end{center}
%\caption{ The ``$\beta$-ray''  images of $^{90}Sr$ source in two energy intervals of  $\beta$-particles.
%Top panel -   $E_{\beta}<1.2~MeV$.
%Bottom panel- $E_{\beta}>1.2~MeV$. }
%\label{xoffline}
%\end{figure}

%Fig.~\ref{bepr}

%Fig.~\ref{bepr}
%Fig.~\ref{edopen}
%%%%%%%%%%%%%%%%%%%%%%%%%%%%%%%%%%%%%%%%%%%%%%%%%%%%%%%%%%%%%%%
%Fig.~\ref{hvdep}
%Fig.~\ref{xdep}
% Fig.~\ref{cosm0504}
%%%%%%%%%%%%%%%%%%%%%%%%%%%%%%%%%%%%%%%%%%%%%
%

\end{document}